\documentstyle[preprint,eqsecnum,aps,tighten]{revtex}
\def\btt#1{{\tt$\backslash$#1}}
\begin{document}
\draft
\preprint{NT@UW-99-22}
\title{
Light Front Nuclear Physics: Toy Models, Static Sources and Tilted Light Front
Coordinates}
\author{P. G. Blunden}
\address{Department of Physics\\University of Manitoba\\
  Winnipeg, Canada R3T 2N2}

\author{M. Burkardt}
\address{Department of Physics\\
New Mexico State University\\
Las Cruces, NM 88003-0001\\U.S.A.}

\author{G. A. Miller}
\address{Department of Physics, Box 351560\\
University of Washington \\
Seattle, WA 98195-1560\\U.S.A.}
\date{\today}
\maketitle
\begin{abstract}
 The principles behind the detailed results of a light-front
mean field theory 
of finite nuclei are
 elucidated by deriving the nucleon mode equation using
 a simple general argument, based on the idea that a static source in
equal time coordinates corresponds to a moving source in light front
coordinates. This idea also allows us to solve
 several simple toy model examples: scalar 
field in a box, 1+1  dimensional bag model,
three-dimensional harmonic oscillator
and the Hulth\'en potential. The latter   provide simplified
versions of  momentum distributions and
form factors of relevance to experiments. In particular, the relativistic
correction to the mean square radius of a nucleus is shown to be very small.
Solving these simple examples suggests another more general 
approach--
the use of
tilted light front coordinates. The simple examples are
 made even simpler.

\end{abstract}

\def\btt#1{{\tt$\backslash$#1}}
\def\be{\begin{equation}}
\def\ee{\end{equation}}
\def\bea{\begin{eqnarray}}
\def\eea{\end{eqnarray}}
\section{Introduction}

Light front techniques\cite{fs}-\cite{harizhang}have long
 been used to analyze high energy experiments
involving nuclear and nucleon targets. In the
 parton model 
 ratio $k^+/p^+$ where  $k^+=k^0 +k^3$ is the plus-momentum of the struck
 quark and $p^+$ is the plus-momentum of the target turns out to be equal to
 the Bjorken variable $x_{Bj}$. Quark distributions
 represent the probability that a quark has a plus-momentum fraction
 $x_{Bj}=k^+/p^+$.
 In nuclear physics we are often concerned with
 the distribution functions which describe the plus-momentum
 carried by 
 the nucleons within the nucleus. Such distributions, which
 depend on $k^+$ of the struck nucleon, are needed to
 analyze deep inelastic lepton-nucleus scattering, and enter also
 in the analysis of high-momentum transfer
 nuclear quasi-elastic reactions such as  $(e,e'), (e,e'p),(p,2p)$.
 If the light front formalism is used, these
 distributions are simply related to the square of the nuclear   ground
 state wave functions. Many nuclear  high momentum transfer 
 are planned, so that it is necessary to derive a relativistic formulation
 which uses $k^+$ variables closely related to experiments and which
 incorporates the full knowledge of nuclear dynamics. 

 If  
 the relevant  nuclear wave functions  depend on $k^+$, the 
 canonical spatial variable  is $x^-=x^0-x^3$. This leaves  
 $x^+=x^0+x^3$ to be  used as a
 time variable, with  the light front Hamiltonian ($x^+$ development operator)
 as $P^-=P^0-P^3$. These are the light front variables of Dirac.
  There are clear  advantages 
in using these
variables,  but
a principal problem arises because the use of $x^+$ as ``time'' and
$x^-,\bbox{x}_\perp$ as ``space'' involves the loss of manifest rotational
invariance. This is especially important in nuclear physics because the
understanding  of magic numbers rests on the $2j+1$ degeneracy of single
particle orbitals of good angular momentum.
Our   previous papers\cite{us} showed how light front techniques
could be used 
to derive a mean field theory for finite-sized nuclei with results that respect
rotational invariance. 

There were three
features that emerged from that  detailed  mean field theory treatment:
\begin{itemize}
\item The variational principle which leads to the field equations involves
   minimizing  the expectation value of $P^-$ subject to the constraint that
   the expectation value of $P^+$ is the same as that of $P^-$.
\item The meson field equations of the light front (LF) formalism are the
    same as those for the equal time (ET) formulation, except for the 
replacement
    $z\to -x^-/2$ in going from the ET to the LF formulation.
    \item The approximate solution of the LF nucleon mode equation, was a phase
      factor times a solution of the ET nucleon mode equation,
      again evaluated with $z\to -x^-/2$. 
  \end{itemize}
The purpose of the present paper is to explain these features in a more
general way and to present some examples of solved problems using light front
techniques. The latter is intended to build up the reader's confidence in the
idea that many different problems can be solved using light front techniques.
The essential feature that allows a simple explanation of the three features is
that in each case
the mode equation involves external potentials  that do not depend on
$x^+$. We show that such  problems are
simply related to  cases of static potentials in the
ET  formulation.
Since using the ET formulation is natural, our
 basic logic  begins by noticing  that
static sources in the usual ET formulation (position  $(x,y,z)$  fixed for all
time $t$) correspond to sources moving with
constant velocity in the LF formulation (because $z+t$ changes with respect to
$z-t$ as $t$ changes. But static  sources depend on the
variable $x^+$  in a manner simple enough to be easily removable with
a simple transformation to a source which is static in light front coordinates.
 The result is a LF theory in which $x^+$ is absent,
and in which the  operator $P^+$ is a part of the $x^+$ development operator. 
The transformation is applied to the Dirac equation in Sect.~II.
The same technique is used to solve four  problems: scalar field in a box,
1+1 dimensional bag model, three dimensional harmonic oscillator and the
Hulth\'en wave function in Sect.~III. We also consider the resulting
light front momentum distribution function as a function of $p^+$ and
$p_\perp$, and also study the related electromagnetic form factors. For the
examples we consider, the only difference between our present approach and
the full calculations in Ref.~\cite{us} is in the support properties.
In the complete theory the wave functions (of positive energy solutions)
have no components with $k^+<0$. This is not kept in the present simple
approach, and
the sizes of  errors introduced depend strongly on the  mass of the
particle involved. For nucleons, the errors are shown to be very small indeed.

The  transformation to a light front static source suggests
the  use of a new set of coordinates--tilted
light front coordinates:  ``time''$ \tau \equiv x^0+x^3 $,  ``space''
$ \zeta \equiv -  x^3 $. With these coordinates
a static source in the ET formulation is also a static source in the
tilted LF formulation, and  the requirement of the constrained  variational
principle that the 
operator $P^+$ be  part of the ``time'' development operator is satisfied
from the very
beginning. 
Tilted  coordinates 
allow us to easily compare LF and ET calculations in
  problems where a single particles moves in an external potential
(fixed source, mean field type problems). 
We introduce these coordinates in Sect.~IV and
demonstrate some utility in Sect.~V
by applying these  to solve  the   simple
problems of Sect.~III. A brief discussion of the results is presented in
Sect.~VI.

In this paper, our use of tilted coordinates is  restricted
 to fixed source, mean field problems  in
which one uses quantum mechanics and not field theory.
This 
is done  to provide the reader with some intuitive ideas about these
coordinates and to illustrate their practical use.
 However, this does by no means exhaust the potential applications
of tilted coordinates. There are many genuine quantum field theory problems
that {\em do} involve also external fields, such as heavy-light 
systems in QCD (e.g. B-mesons) or a QFT treatment of large nuclei.
For such systems, the relation between ET and  LF is more complicated than
the situations discussed here
because the microscopic degrees of freedom (the quanta)
in terms of which the theory is formulated are  different in these two 
approaches and the LF approach {\em can} provide new insights
into observables that probe light-like correlations.
The use of tilted coordinates is that they provide a significant shortcut
towards constructing the LF Hamiltonian.  

\section{Alternate Derivation of the Nucleon Mode Equation}

In Ref.~\cite{us} the constrained minimization (of $P^-$)
led to a new equation for
the single-nucleon
modes (Eq.~(4) in the short paper, (3.28) of the long paper).  
In order to illuminate the physics we present another, more heuristic,
derivation of the nucleon single particle wave equation.
The first step towards formulating the
mean field approximation in light front LF coordinates is to
develop the
formalism for static potentials (static in a rest frame!) on the LF.
For this purpose,
we start from the Dirac equation for a static potential in normal coordinates
\begin{equation}
\left[i\gamma^\mu\partial_\mu - m - g_SV_S({\vec r}) - g_V
\gamma^0V_0 ({\vec r})  \right] \psi' =0.
\label{eq:dirac_et}
\end{equation}
Since $\gamma^0= (\gamma^++\gamma^-)/{2}$ and since couplings
using  the ``bad'' component $\gamma^-$ are difficult to handle in
the LF framework, we perform the transformation 
\begin{equation}
\psi^\prime = e^{ig_V \Gamma} \tilde{\psi}^\prime,\label{tg}
\end{equation}
where
\begin{equation}
\partial_3 \Gamma = V_0 ,
\end{equation}
and   $\Gamma$ does not depend on time.
Using the transformation (\ref{tg}),
one  finds that $\tilde{\psi}^\prime$ couples to the vector field only
via $\gamma^0+\gamma^3$ (the LF-$\gamma^+$ component!)
\bea
0 &=& \left[i\gamma^\mu\partial_\mu
+i{\vec \gamma}_\perp\cdot ( {\vec \partial}_\perp\Gamma)
  - m - g_SV_S({\vec r}) - g_V
\left(\gamma^0+\gamma^3\right) V_0 ({\vec r})  \right]\tilde{\psi}^\prime .
\label{psip}\eea

Simply rewriting Eq. (\ref{eq:dirac_et}) in LF coordinates, with
$\gamma^\pm \equiv (\gamma^0 \pm \gamma^3)$ and 
$x^\pm \equiv (x^0\pm x^3)$, yields 
\begin{eqnarray}
& &\left[{1\over2}\left(i\gamma^+\partial^- + i\gamma^-\partial^+\right)
+i{\vec \gamma}_\perp\cdot\left( {\vec \partial}_\perp
  + g_v {\vec \partial}_\perp
  \Gamma({\vec r}_\perp, \frac{x^+-x^-}{{2}}) \right)- m \right.
\label{eq:dirac_lf1}
\\
& &\quad \left.- 
V_S({\vec r}_\perp, \frac{x^+-x^-}{{2}} )- 
\gamma^+V_0 ({\vec r}_\perp, \frac{x^+-x^-}{{2}} )
\right] \tilde{\psi}^\prime =0. \nonumber\end{eqnarray}

Even though the potential is static in the equal time formulation,
the Dirac equation for the same potential in light-front coordinates
is LF-``time'', i.e. $x^+$, dependent.
The physical origin for this result is that a static source in a 
rest-frame corresponds to a uniformly moving source on the 
light-front. Given that the time dependence of the external fields is
only due to a uniform translation, it should be easy to transform 
Eq. (\ref{eq:dirac_lf1}) into
a form which contains only static (with respect to $x^+$)
potentials. For this purpose, we consider the equation of motion 
satisfied by Dirac fields which are obtained by an $x^+$ (LF-time) 
dependent translation 
\begin{eqnarray}
\tilde{\psi}^\prime  ({\vec x}_\perp,x^-,x^+)
&\equiv& e^{-ix^+P^+/2}{ \psi} ({\vec x}_\perp,
x^-,x^+) 
\label{eq:tilde}
\end{eqnarray}
Using $P^+=-i2{\partial\over \partial x^-}$, we find
\begin{eqnarray}
e^{ix^+ P^+/2} f\left({x^+-x^-\over2}\right) e^{-ix^+ P^+/2} &=&
f\left({-x^-\over 2}\right)
\nonumber\\
e^{ix^+ P^+/2} \partial^-e^{-ix^+ P^+/2} &=& 
\partial^- - \partial^+ \label{trans}
\end{eqnarray}
so that the equation of motion for $\psi$ takes the form:
\begin{eqnarray}
& &[{1\over 2}i\gamma^+(\partial^--\partial^+) +{1\over2}
  i\gamma^-\partial^+
+i{\vec \gamma}_\perp\cdot
({\vec \partial}_\perp+ig_v\Gamma({\vec r}_\perp,-\frac{x^-}{{2}}))-m 
\label{eq:dirac_lf2}
\\
& &\quad - 
g_SV_S({\vec r}_\perp, -\frac{x^-}{{2}} )- 
g_V\gamma^+V^- ({\vec r}_\perp, -\frac{x^-}{2} )
] {\psi} =0, \nonumber\end{eqnarray}
with $V^-=V_0$.
The translated fields satisfy an equation of motion 
with  potentials that do not depend on $x^+$. Moreover, the 
static potentials evaluated at ${\vec r}$ correspond to light front potentials
evaluated at $({\vec r}_\perp,-\frac{x^-}{{2}})$.
A simple derivation of this can be obtained from evaluating $z=(x^+-x^-)/2$
at $x^+=0$.

That the  result (\ref{eq:dirac_lf2}) is  the same as the equations for
$\psi_\pm$ in Ref.~\cite{us}.  
 can be seen by making a decomposition  
 into a dynamical and a constraint equation.
Multiplication of Eq. (\ref{eq:dirac_lf2}) by  $\gamma^+$ from the left
yields
a constraint equation (no $x^+$-derivative !)
\begin{equation}
i \partial^+ \psi_-
=  \left[
i{\vec \alpha}_\perp\cdot ({\vec \partial}_\perp+ig_V
({\vec \partial}_\perp\Gamma))+\beta m  +V_S
\right] \psi_+ 
\label{eq:tconstr}
\end{equation}
where, as usual, $\psi_\pm \equiv \frac{1}{2} \gamma^0
\gamma^\pm \psi$.
Multiplication of Eq. (\ref{eq:dirac_lf2}) by  $\gamma^-$ from the left
yields
an equation for $\psi_+$:\begin{equation}
i (\partial^--\partial^+-ig_vV^-) \psi_+
=  \left[
i{\vec \alpha}_\perp\cdot ({\vec \partial}_\perp+ig_V
({\vec \partial}_\perp\Gamma))+\beta m  +V_S
\right] \psi_-.
\label{eq:dyn}
\end{equation}
One may use 
 the constraint equation (\ref{eq:tconstr}) to eliminate 
$\psi_{-}$ in Eq. (\ref{eq:dyn}) to obtain the  
 equation of motion for 
for the dynamical degrees of freedoms. The results (\ref{eq:tconstr})
and (\ref{eq:dyn}) are the desired equations.

\section{Simple problems}
In order to illustrate the application of the LF formalism for
static external potentials, let us consider a few simple examples:
a scalar field in a box with vanishing boundary conditions,
a Dirac field in a ``bag'' with bag boundary conditions, 
the 3 dimensional (scalar) harmonic oscillator, and the 3 dimensional Hulth\'en
potential.

\subsection{Scalar Field in a Box}

Let us first consider a box of length $L$ (extending from $0$ to $L$)
with vanishing boundary conditions  and determine the eigenstates
in the equal time framework. The two degenerate solutions to the
free Klein-Gordon equation are left and right moving plane waves
$
\phi_\pm(x,t)=e^{-iE t} e^{\pm ikz}
$
where $E=\sqrt{k^2+m^2}$. Using the boundary conditions to match 
coefficients and wave number in the superposition of these two 
solutions one obtains the familiar solution
\be
\phi_n (z,t) = e^{-iE_n t} \sin (k_nz), 
\ee
where
\bea
k_n &=& \frac{n\pi}{L}\quad \quad \quad n=1,2,3,... \nonumber\\
E_n^2 &=& m^2 + k_n^2 .
\label{eq:En}
\eea

In a light front calculation the boundary conditions
depend on $x^+$ 
because $z=(x^+-x^-)/2$.
This dependence can be eliminated using transformation of the form
of (\ref{eq:tilde}). We start with the Klein-Gordon equation 
\be
(\partial^+\partial^-+m^2) \tilde{\phi}(x^-,x^+)=0
,\ee
and make the transformation 
\be \tilde{\phi}=e^{-ix^+P^+/2}\phi\ee
to find 
\be
\left(\partial^+(\partial^--\partial^+)+m^2\right)\phi(x^-,x^+)=0,
\ee
with (according to the translation \ref{trans}) 
 the boundary condition 
\be \phi(x^-=0,-2L,x^+)=0.\label{bc1}\ee
We find solutions of the form
\be \phi(x^-,x^+)=e^{-ik^-x^+/2} e^{-i k^+x^-/2}, \ee
with 
 the
dispersion relation:
\be
k^- = \frac{m^2}{k^+} + k^+
\label{eq:kab1}.
\ee
Before taking the boundary condition into account, there are two
linearly independent solutions $e^{-ik_a^+x^-/2}$ and 
$e^{-ik_b^+x^-/2}$, each of which has the  value of $k^-$ provided by
Eq.~(\ref{eq:kab1}).
Imposing the boundary condition at $x^-=0$ leads to the form:
\be
\phi(x^+,x^-)=e^{-ik^-x^+/2}
\left(e^{i{k_a^+x^-/ 2}}-e^{i{k_b^+x^-/ 2}} \right)\label{lftw}
.\ee
The vanishing of $\phi$ at the other boundary ($x^-=-2L$) implies
\be
k_a^+ -k_b^+ = n \frac{2\pi}{L}  
\quad \quad \quad \quad n=1,2,...
\label{eq:bc2}
\ee
which constrains the allowed energy eigenvalues $k_n^-$. Using
Eq.~(\ref{eq:kab1}) for $k^-_a=k^-_b=k^-_n$ shows that
 the two independent solutions to  Eq. 
(\ref{eq:kab1}) are related by
\be
k_b^+ = \frac{m^2}{k_a^+}
\ee
and thus the quantization condition for the LF momenta in a
stationary box can be written as
$k_{a,n}^+ -\frac{m^2}{k_{a,n}^+} = n \frac{2\pi}{L}$.
Hence the quantized energies satisfy
\bea
{k_n^-}^2 &\equiv& \left(\frac{m^2}{k^+} + k^+\right)^2
=\left(\frac{m^2}{k^+} - k^+\right)^2 + 4m^2
\nonumber\\
&=& \left(n\frac{2\pi}{L}\right)^2+4m^2
= 4\left[\left(n \frac{\pi}{L}\right)^2+m^2
\right] ,
\label{eq:kn-}
\eea
which is consistent with Eq. (\ref{eq:En}) because $k_n^-$ is  identified
 with ${2}E_n$. 

The fact that the  LF-ansatz
Eq. (\ref{lftw}) consists of two waves ---
both moving in the same direction (both LF-momenta positive!) ---
has a very intuitive interpretation.
In ordinary coordinates, the stationary solutions for a particle in a 
box with hard walls is described by a superposition of plane waves moving
in opposite directions. Since the LF momentum of a plane wave is given by the
sum of the particle's momentum in the z-direction and its energy,
the two waves moving in opposite directions will have different
LF-momenta. However, both LF momenta are positive, since the ET energy is 
larger than the absolute value of the momentum. 
\subsection{1+1 Dimensional Bag Model}
As a second  example, let us consider the 
bag model. For simplicity, we will only consider the 1+1 dimensional
case here.
In 1+1 dimensions, Dirac matrices are $2\times 2$ matrices and we choose to
work in the chiral representation
\bea
\gamma_0 = \left( \begin{array}{cc} 0 & 1 \\ 1 & 0 \end{array} \right)
\quad \quad \quad \quad \quad \quad 
\gamma_1 = \left( \begin{array}{cc} 0 & 1 \\ -1 & 0 \end{array} \right) .
\label{eq:gammaET}
\eea
In order to solve the Dirac equation for a ``bag''
\bea
i\left(\gamma_0 \partial_t - \gamma_1 \partial_x\right) \psi &=& m \psi
\quad \quad \quad \quad (\mbox{inside})\nonumber\\
n^\mu \gamma_\mu \psi &=& i \psi
\quad \quad \quad \quad (\mbox{at boundary}),
\label{dirac}\eea
where $n^\mu$ is normal to the surface,
we make a stationary wave ansatz
\be
\psi = \left( \begin{array}{c} \chi_1 \\ \chi_2 \end{array} \right)
e^{-iEt},
\ee
yielding
\bea
E \chi_2 - i \chi_2^\prime &=& m \chi_1 \nonumber\\
E \chi_1 + i \chi_1^\prime &=& m \chi_2 \quad \quad \quad \quad (\mbox{inside})
\nonumber\\
\chi_2 &=& i\chi_1 \quad \quad \quad \quad \quad(x=L)
\nonumber\\
\chi_2 &=& -i\chi_1 \quad \quad \quad \quad (x=0).
\label{eq:bag1}
\eea
The case $m=0$ is particularly easy to solve, and one finds
\bea
\chi_1^n &=& e^{-iE_nx} \nonumber\\
\chi_2^n &=& -ie^{iE_nx} \nonumber\\
E_n &=& \left( n+ \frac{1}{2} \right) \frac{\pi}{L} 
\quad \quad \quad \quad
n=0,1,...
\label{eq:bagEn}
\eea

Turn  now to the light front calculation.
We use the same representation for the $\gamma$-matrices 
as above (\ref{eq:gammaET}), yielding 
\bea
\gamma^+ = \left( \begin{array}{cc} 0 & 0 \\ {2} & 0 \end{array} \right)
\quad \quad \quad \quad \quad \quad 
\gamma^- = \left( \begin{array}{cc} 0 & {2} \\ 0 & 0 \end{array} \right) .
\label{eq:gammaLF}
\eea
Our version of the Dirac equation is given in  1+1 dimensions (after the
transformation (\ref{eq:tilde})) leads to the 1+1 dimensional version of
 (\ref{eq:dirac_lf2}) ) as
\bea
i\partial^-\psi_+&=&i\partial^+\psi_++m\psi_- \nonumber\\
i\partial^+\psi_-&=&m\psi_+
.\label{lfdd}\eea
We look for a solution of the form:
\be
\psi(x)=e^{-ip^-x^+/2}\psi(x^-),
\ee
which gives for (\ref{lfdd}):
\bea p^-\psi_+&=&i\partial^+\psi_++m\psi_-\nonumber\\
i\partial^+\psi_-&=&m\psi_+
,\label{lfde}\eea
and therefore
\be
 p^-\psi_+=\left({m^2\over i\partial^+}+i\partial^+\right)\psi_+.\ee
Using plane wave solutions of the form $\psi_+=e^{-ik^+_ax^-/2}$ or
$\psi_+=e^{-ik^+_bx^-/2}$, leads immediately to
$
p^-={m^2\over k_{a,b}^+} +k^+_{a,b},$
so that once again $ k^+_b=m^2/k_a^+$. The boundary conditions of
Eq.~(\ref{dirac}) are given in light front
coordinates as
\bea
\psi_-&=&i\psi_+\quad\quad (x^-=-2L)\nonumber\\
\psi_-&=&-i\psi_+\quad\quad (x^-=0)
\label{mbc}.\eea
For $m=0$ the solutions to Eqs.~(\ref{lfde}) are given by
\bea
\psi_+&=&e^{-ip^-x^-/2}\nonumber\\
\psi_-&=&const, 
\eea
so the boundary conditions (\ref{mbc})
lead immediately
to
\be
p_n^- = \left( n+\frac{1}{2} \right) \frac{2\pi}{L},
\ee
which is $2E_n$.

\subsection{Three Dimensional Harmonic Oscillator}
In ordinary coordinates, the Klein-Gordon  equation for a 3 dimensional
relativistic oscillator
\be
E^2 \phi  = \left[{\vec p}^2 + m^2 + \kappa {\vec x}^2 \right] \phi
\ee
closely resembles the equation for a nonrelativistic harmonic 
oscillator and we can thus immediately write down its energy eigenvalues
\be
E_n^2 = m^2 + \omega \left( n+ \frac{3}{2} \right),
\label{eig}\ee
where $\omega^2 = 4 \kappa$.

The light front version of the Helmholtz equation is
\be
\left(\partial^-\partial^+-\nabla^2_\perp+m^2+\kappa
(x^2_\perp+(x^+-x^-)^2/4)\right)\phi=0.\ee
Once again we see a familiar pattern. The static potential in ET coordinates 
becomes a `time'' or $x^+$ dependent potential in LF coordinates. Once again
this dependence is of a simple form; it can be transformed away using
\be
\phi=e^{-i{x^+P^+\over2}}\chi
.\ee
The use of Eqns.~(\ref{eq:tilde})    and (\ref{trans})
then leads to the result
\be 
\left((\partial^--\partial^+)\partial^+-\nabla^2_\perp+m^2+\kappa (x^2_\perp+
({x^-\over 2})^2)\right)\chi=0 .\ee In the absence of interactions the 
$P^-$ operator would consist of the usual term plus an  additional $p^+$ 
operator.
We look for a solution of the form:
\be
\chi(x)=e^{-ip_n^-x^+/2}\chi(x^-,\vec {x}_\perp).\ee
This, along with completing the square,  leads  to  the result
\be
\left( -(\partial^++ip^-_n/2)^2-\nabla^2_\perp+m^2+\kappa (x^2_\perp+
({x^-\over 2})^2\right)\chi=(p^-_n/2)^2\chi .\ee
One converts the operator $ (\partial^++ip^-_n/2)^2$ to $(\partial^+)^2$ using
yet  another transformation:
\be
\chi(x^-,\vec{x}_\perp)=e^{-ip_n^-x^-/4}F(x^-,\vec{x}_\perp)\ee
to find
\be \left( -(\partial^+)^2-\nabla^2_\perp+m^2 +\kappa (x^2_\perp+
({x^-\over 2})^2\right)F=(p_n^-/2)^2 F.\ee
This is the same form as 
the equation in the equal time coordinates and $p^-_n/2$
takes on the values of $E_n$.

     We seem to be getting the same results as  in the ET development. So one 
might wonder why we are doing the light front at all. The point is that we are 
      able to compute the light front
wave functions that depend on $x^-$, or in  momentum space depend on $p^+$.
The wave functions of the ground  state is
     given by
     \be
  \chi_0(x^-,\vec{x}_\perp)=e^{ip_0^-x^-/4}N_0 \exp
  {\left(-{1\over2}\sqrt{\kappa}
    (x_\perp^2+{{x^-}^2\over
          4})\right)}.
      \ee

      The number density $n_0(p_\perp,p^+)$  is  defined as the square
      of the momentum space version of $\chi_0$.
This quantity is accessible in high energy proton and electron nuclear
quasi-elastic reactions.
      It is useful to define
      the light front variable \be\alpha\equiv p^+/({p_0^-\over2}).
      \label{lfv}\ee 
Then one may
      easily determine that
      \be n_0(p_\perp,\alpha)=\tilde{N}_0e^{-{p_\perp^2\over\sqrt{\kappa}}}
      e^{{-(p_0^-(\alpha-1))^2\over4\sqrt{\kappa}}}.
      \label{nho}\ee
Note that one finds the same $p_\perp$ distribution for each value of the
variable $\alpha$.  This is not a general 
feature of light-front wave functions, as we show in the next sub-section.

In an exact calculation, the number density should vanish for values of
$\alpha$ that are not between 0 and 1. This is referred to as the 
 support, which we examine here.   
How large can the value of $\alpha$ be?
For large values of the particles mass 
$m$, ($m\gg\kappa^{1/4})$ (which corresponds to
the situation of
nuclear physics in which the 
product of the nucleon mass and the nuclear radius is a very large number)  
the value of $\alpha$ must always be close to unity. If $m\to 0$ then
the behavior is
$\sim e^{-3(\alpha-1)^2}$. We may better understand this factor 
by considering an extremely simple model--  the nucleon consists of three
 massless   quarks moving in an harmonic oscillator potential. We shall
 compute the quark distribution 
function $q(x)$, which here is the same as $n_0$ integrated  over
$p_\perp$  and evaluated as a  function of the Bjorken variable, $x$.
For massless quarks, Eq.~(\ref{eig}) gives
\be ({p_0^-\over 2})^2=3\sqrt{\kappa}.\ee Furthermore,
  in the target rest frame  $P_N^+$ is 
the mass of the nucleon 
$3{p_0^-\over2}$. Thus using Eq.~(\ref{lfv})
$\alpha=3{p^+/P_N^+}=3x$. The last equality is from
the parton model in which the ratio of the quark and target plus momenta is 
${Q^2\over 2m_N\nu}=x$. Thus,  we find
\be
q(x)\propto e^{-27(x-1/3)^2}.\ee
In this limit of massless quarks, the value of $x$ (in this un-evolved)
quark distribution function is constrained to be very close to $1/3$. One would
 naively expect the value of $x$  to easily exceed unity, since we have
used a mean field model for a three-quark system. This does not occur. If $x=1$
the value of the $q$ is $e^{-12}$ its peak value, which is reasonably
small. If $x$ approaches $0$,
one finds a factor $e^{-3}$, instead of the required $0$ so 
there is an inherent inaccuracy of some 5\%. 
If one included a
non-zero value of the quark mass, the support properties would
be improved because 
\be
q(x)\propto e^{-({9m^2\over\sqrt{\kappa}}+27)(x-1/3)^2}.\ee
In constituent quark models $m^2=\sqrt{\kappa}$, so that at $x=0$ one finds
a factor of $e^{-4}$. In nuclear physics ${m^2\over \sqrt{\kappa}}$ is
of the order of $(5R_A/ \rm{Fm})^2$ which taking $R_A=4$ Fm, yields
$e^{- 400}$ at $x=0$. In that case, there is no problem with the support.

It is of interest to compute the electromagnetic form factor of the ground
state. This quantity has  been often  measured in elastic electron scattering,
and
the sizes of nuclei have been determined as one of the classic achievements of
nuclear physics. Interest in this topic has been revised because of a
recent proposal to Jefferson Laboratory to use parity-violating electron
scattering to measure the neutron radius\cite{happex}.
A high precision is needed and
can be obtained provided one knows the proton distribution. 
Therefore one 
needs to examine the influence of  small effects such as  relativistic
corrections.
One works using a reference frame in which the plus component of the four
vector $q^\mu$ of the virtual photon vanishes, so that $Q^2=-q^2=q^2_\perp$.
In this case the form factor $F(Q^2)$
(matrix element of the plus component of the
electromagnetic current operator) is given by
\be F(Q^2)=\int d^2p_\perp d\alpha\; \chi_0(\alpha,\bbox{p}_\perp)
 \chi_0(\alpha,\bbox{p}_\perp-\alpha \bbox{q}_\perp),
 \ee
in which the influence of relativity appears in the integral over $\alpha$
and the factor $\alpha$. For the harmonic oscillator ground state we
find:
\be F(Q^2)= N\int d^2p_\perp d\alpha\; e^{-\left(
    {p_\perp^2\over \sqrt{\kappa}}
    +{q_\perp^2\alpha^2\over 4\sqrt{\kappa}}\right)}
e^{-{{p_0^-}^2\over 4\sqrt{\kappa}}(\alpha-1)^2}\label{form}
,\ee
where \be {{p_0^-}^2\over 4}=m^2+3\sqrt{\kappa}.\ee

Our purpose here is the study of nuclear physics, so we are interested in
the non-relativistic limit and the corrections to it.
To this end, we define a variable $p_z$ using
\be \alpha=1+{p_z\over m}.\ee The non-relativistic limit of (\ref{form}) is
obtained by letting $m$ approach infinity. Then $ {{p_0^-}^2\over 4}=m^2$ and
we find
\bea F_{NR}(Q^2)&=& N_{NR}\int {d^3p\over m} e^{-\left(
    {p_\perp^2\over \sqrt{\kappa}}+{q_\perp^2\over
      4\sqrt{\kappa}}\right)}\nonumber\\
&=& e^{-{Q^2\over 4\sqrt{\kappa}}}
.\eea
The mean square radius  $-6{d F(Q^2)\over d Q^2}\mid_{Q^2=0}$, is given by
\be R^2_{NR}={3\over 2} {1\over\sqrt{\kappa}}.\ee

The leading corrections to this will be of order
$p_z^2/m^2\sim \sqrt{\kappa}/m^2$. We define a semi-relativistic limit $SR$ via
the use of (\ref{form}) and keeping the leading correction terms. Performing
the straightforward evaluations leads to the result:
\be
\delta\equiv{R_{SR}^2-R_{NR}^2\over R_{NR}^2}={3\over 2}\sqrt{\pi}
{\sqrt{\kappa}\over m^2},
\ee
or
\be \delta\approx{\sqrt{\pi}\over 12 A^{2/3}}.\ee
This corresponds to very small (0.004) effects for large nuclei $A\sim 200.$

\subsection {Light Front Hulth\'en Wave Function}
\begin{figure}
\unitlength1cm
\begin{picture}(10,8)(-3,2.5)
\includegraphics{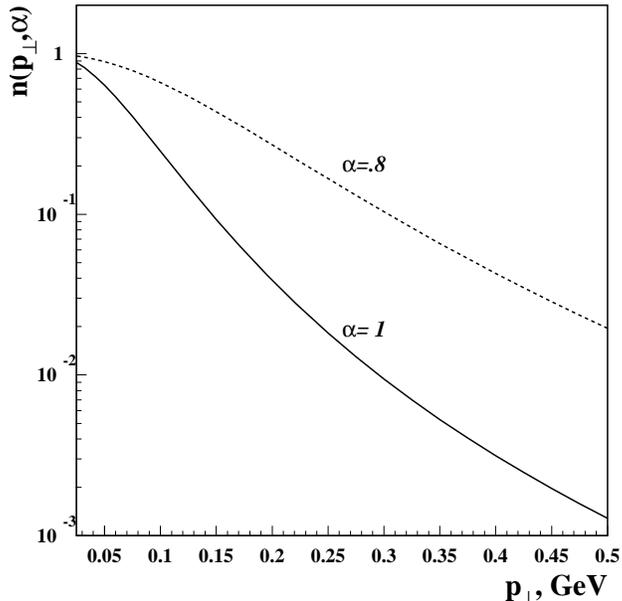}
\end{picture}
\caption{Light front momentum distribution as a function
  of $\alpha$ and $p_\perp$.
 }
\label{fig:hulth}
\end{figure}

Any static potential of the form $V(\vec{x}^2=x_\perp^2+z^2)$
 can be solved on the light front. The transformation
 (\ref{eq:tilde}) corresponds to including the $p^+$ term in the
 $x^+$ development operator and 
a  simple prescription of replacing $z$ by $-x^-/2$ in $V$.
We present here  the solution for the Hulth\'en potential. This allows us to
demonstrate an
interesting contrast between the implications of different forms of 
potentials.
In the equal time formulation we have the wave equation:
\be
E^2 \phi  = \left[{\vec p}^2 + m^2 + V^H ({\vec x}^2) \right] \phi,
\ee
in which $V^H$ is chosen so that the lowest energy solution is 
\be
\phi(r)=N(e^{-ar}-e^{-br}),
\ee
where $b>a$. The eigenenergy is given by
\be
E=\sqrt{m^2-a^2}.
\label{EH}\ee
The light front version of the wave equation is
 \be
    {{p_n^-}^2\over 4}\chi(x^-,\vec{x}_\perp)=\left[\left(-2i{\partial\over
          \partial x^-}\right)^2+p_\perp^2+V^H\left(x_\perp^2+{{x^-}^2\over
          4}\right)\right]\chi(x^-,\vec{x}_\perp)
     . \ee The lowest value of $p_n^-/2$ is clearly the
     same as $E$ of Eq.~(\ref{EH}), and the
     wave function is
     given by
     \be
  \chi_0(x^-,\vec{x}_\perp)=e^{ip_n^-x^-/4}N_0^H
  \left[\exp{\left(-a\sqrt{x_\perp^2+{{x^-}^2\over 4}}\right)}
    -\exp{\left(-b\sqrt{x_\perp^2+{{x^-}^2\over 4}}\right)}\right]
  .
      \ee
The momentum distribution $n^H_0(p_\perp,p^+)$ obtained here provides an
interesting contrast with that of the harmonic oscillator (\ref{nho}).
We find
\be
n^H_0(p_\perp)=\tilde{N}_0^H\left[{1\over \left( a^2+p_\perp^2+(p^-_n/2)^2
    (\alpha-1)^2 \right)^2}
-{1\over \left( (b^2+p_\perp^2+(p^-_n/2)^2
    (\alpha-1)^2\right)^2}\right]
.\ee
It is clear that one finds a different $p_\perp$ distribution
for each value of
$\alpha$. The distribution is a broader function of $p_\perp$ for larger values
of $\alpha$; see the Fig.~\ref{fig:hulth}. The results in the figure
are obtained using $m$=.94 GeV, $E$=.932 GeV and $b=5 a$.
An experimental hint of such a behavior has been found
recently\cite{ep}.
\section{Tilted Light-Front Coordinates}

We have worked 
out several examples involving  potentials that are independent of time in
equal time coordinates.
This leads to an $x^+$-dependent interaction when light front coordinates are 
used.  However  this dependence is very simple, and   
  in each case we have removed this by using 
a version of Eqs.~(\ref{eq:tilde}). 
In each case  we have derived a light front wave equation in which the  
 kinetic energy 
has the same form as the
standard LF-kinetic energy plus an additional term linear in the 
momentum which is identical to the recoil term in the static
source formalism on the LF!  It is  worthwhile to see if 
there is a more general way to remove this dependence, once and for all,
 by finding a set
of coordinates in which static sources in equal time coordinates are  
also described by static sources in a coordinate system that is very 
much like that of the light front.

\begin{figure}
\unitlength1cm
\begin{picture}(15,8)(-.7,0)
\includegraphics{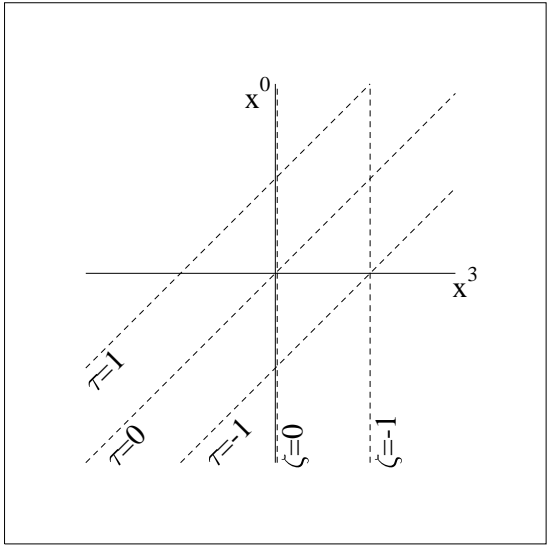}
\end{picture}
\caption{Comparison between lines of equal $x^0$ and $x^3$ (full lines)
and lines of equal $\tau$ and $\zeta$ (dashed).}
\label{fig:tilt}
\end{figure}

For this purpose, we introduce new ``tilted'' LF-coordinates
\begin{eqnarray}
\tau &\equiv& x^0+x^3 
\nonumber \\
\zeta &\equiv& -  x^3  
\label{eq:tilt},
\end{eqnarray}
i.e.
\begin{eqnarray}
x^0 &=& 
\tau + \zeta 
\nonumber\\
x^3 &=& - \zeta 
\label{eq:itilt}
\end{eqnarray}
with ${\vec x}_\perp$ as usual.
These coordinates very much resemble LF-coordinates, since
(Fig. \ref{fig:tilt})
\begin{itemize}
\item surfaces of constant $\tau$ are the usual LF hypersurfaces
and thus quantization is very similar to LF quantization.
\item correlation functions at $\tau=0$ in the $\zeta$ direction
yield the usual LF distributions. Therefore, a lot of the familiar
LF-phenomenology (e.g. structure functions and wave functions) 
can still be used.
\end{itemize}
However, in contradistinction to light-front coordinates, static
sources in a rest-frame are also described by static sources in 
above tilted coordinates, since
\begin{eqnarray}
\partial_0 &=& 
\partial_\tau \nonumber\\
\partial_3 &=& \partial_\tau 
- 
\partial_\zeta ,
\label{eq:partial}
\end{eqnarray}
i.e. if $\partial_0V=0$ then $\partial_\tau V=0$.
For a static ($\partial_tV=0$) potential, one thus finds for the
static potential ${\cal V}({\vec x}_\perp, x^3)$ 
\begin{equation}
{\cal V}({\vec x}_\perp,x^3) =
V({\vec x}_\perp, -\zeta) 
\label{eq:calV}
\end{equation}

Furthermore, using Eq. (\ref{eq:partial}), one finds 
for the longitudinal part of the kinetic energy operator 
\begin{equation}
\partial_0^2-\partial_{3}^2 =2\partial_\tau \partial_\zeta
- \partial_\zeta^2 .
\label{eq:kinetic}
\end{equation}
The Lagrangian for a scalar field interacting with  an external field
can then be
written as 
\bea
{\cal L} &=& \frac{1}{2}\left[
\partial_\mu \phi \partial^\mu \phi -\left(m^2+V({\vec x})\right) \phi^2\right]
\label{eq:Ltiltscalar}
\\ &=& \partial_\tau \phi \partial_\zeta \phi
-  \frac{1}{2}\left(\partial_\zeta \phi\right)^2 
- \frac{1}{2}\left({\vec \partial}_\perp \phi \right)^2
- \frac{m^2+V({\vec x})}{2} \phi^2 \nonumber
\eea
in tilted coordinates, the canonical momenta are the longitudinal ``space''-derivatives
of the fields 
\begin{equation}
\Pi \equiv \frac{\partial {\cal L}}{\partial (\partial_\tau \phi )
}= \partial_\zeta \phi \equiv \partial_-\phi
\end{equation}
--- just like in genuine LF-coordinates.
This should not come as a surprise, since the constant ``time''
surfaces in our tilted coordinates and in LF coordinates are
identical (Fig. \ref{fig:tilt}).

The kinetic energy operator differs from the one
in LF-coordinates. Using Eq.(\ref{eq:kinetic})
in the form $2{\cal E}{\cal P}-{{\cal P}^2-\vec P}_\perp^2-m^2=0$, where
${\cal E} $ is conjugate to $\tau$ and ${\cal P}$ is conjugate to $\zeta$,
one finds  the kinetic energy
of a free particle in tilted coordinates:
\begin{equation}
{\cal E} = \frac
{m^2+{\vec P}_\perp^2 }{ {\cal P}} +
{\cal P}.
\label{eq:Etilted}
\end{equation}
For non-interacting systems,
this  is identical to the expression derived in LF-coordinates
including the  ``Lagrange multiplier''\cite{us}.

The point of using the tilted LF coordinates is that the term ${\cal P}$ is
included automatically. The energies $\cal E$ obtained using
tilted LF coordinates are the same as $P^-$ obtained in the usual LF
dynamics including the Lagrange multiplier term. We expect that 
 the ${\cal E}$ of tilted LF must be the same as the $P^-$ of LF because
both formulations use the same quantization (i.e. equal 'time') hypersurfaces.
The only place where they differ is the space  direction, where the
difference between the  spatial coordinates $x^-$ and $\zeta$ is a shift $x^0$.
This is a simple translation --- hence the term linear in the momentum in
the tilted coordinates Hamiltonian. However, it turns out that in the
mean field approach one needs to introduce a Lagrange multiplier term to
fix the momentum, and the two Hamiltonians have an identical
form.

We shall see how this works out by solving the 
same  simple problems as above using
the tilted  LF dynamics.

\section{Simple Problems on the Tilted Front}
In order to illustrate the application of the LF formalism for
static external potentials, let us consider a few simple examples:
a scalar field in a box with vanishing boundary conditions,
the 3 dimensional (scalar) harmonic oscillator and
a Dirac field in a ``bag'' with bag boundary conditions.

\subsection{Scalar field in a box}
In order to derive the tilted LF solution for this example, we first note
that vanishing boundary conditions are frame independent, i.e. a
stationary box with vanishing boundary condition in an equal time
framework corresponds also in tilted LF coordinates to a stationary
box with vanishing boundary condition. 
Before taking the boundary condition into account, there are two
linearly independent solutions $e^{-ik_a^+\zeta}$ and 
$e^{-ik_b^+\zeta}$
for a given energy, which is the same according to Eqs.~(\ref{eq:Etilted}) 
as that of Eq.~(\ref{eq:kab1}).

The only superposition which satisfies the requirement that the 
wave function vanishes at $\zeta=0$ is given by
\be
\phi(\zeta)=e^{-ik_a^+\zeta}-e^{-ik_b^+\zeta} .
\label{eq:2waves}
\ee
Vanishing of $\phi(x^-)$ at the other boundary ($\zeta=-L$) implies
\be
k_a^+ -k_b^+ = n \frac{2\pi}{L}  
\quad \quad \quad \quad n=1,2,...
\label{eq:bc1}
\ee
which constrains the allowed energy eigenvalues $k_n^-$. Using
Eq.~(\ref{eq:Etilted}) for $k^-_a=k^-_b=k^-_n$ shows that
 the two independent solutions to  Eq. 
(\ref{eq:kab1}) are related by
\be
k_b^+ = \frac{m^2}{k_a^+}
\ee
and thus the quantization condition for the LF momenta in a
stationary box can be written as
$k_{a,n}^+ -\frac{m^2}{k_{a,n}^+} = n \frac{2\pi}{L}$.
Hence the quantized energies satisfy Eq.~(\ref{eq:kn-}),
which is consistent with Eq. (\ref{eq:En}) provided one identifies
$k_n^-$ with ${2}E_n$ (\ref{eq:partial}). 
\subsection{1+1 Dimensional Bag Model}
Making a stationary wave ansatz  
\be
\psi = \left( \begin{array}{c} \psi_{+} \\ \psi_{-} \end{array} \right)
e^{-i{\cal E}\tau/2},
\ee
one thus writes  the equation of motion for
a free Dirac particle (\ref{dirac}), in tilted coordinates by multiplication
by either $\gamma^+$ or $\gamma^-$. This gives
\bea
{\cal E} \psi_{+} - i\partial_\zeta\psi_{+}  
&=& m\psi_{-} 
\nonumber\\
 - i\partial_\zeta \psi_{-} &=& m \psi_{+}, 
\label{eq:lfbag}
\eea
where $\partial_\zeta\equiv {\partial\over \partial\zeta}$.
We  use the second of Eqs.~(\ref{eq:lfbag}) in the first to 
to obtain  the ``dynamical'' component:
\be
{\cal E} \psi_{+} = \left[\frac{m^2}{i\partial_\zeta}
+ i\partial_\zeta \right] \psi_{+} .
\ee
As is the case for free scalars, one finds in general two linearly independent
plane wave solutions for each energy $\psi_{+} = e^{-ik_a^+ \zeta}$ and
$\psi_{+} = e^{-ik_b^+ \zeta}$ where ${\cal E} = \frac{m^2}{k_{a,b}^+}
+k_{a,b}^+$ and $k_b^+= \frac{m^2}{k_{a}^+}$.
Note that the boundary condition (\ref{dirac}) mixes $\psi_{+}$ and 
$\psi_{-}$
\bea
\psi_{-}&=&i\psi_{+} \quad \quad \quad \quad (\zeta = -L)
\nonumber\\
\psi_{-}&=&-i\psi_{+} \quad \quad \quad \quad (\zeta = 0) .
\label{eq:lfbc}
\eea
This mixing should not come as a surprise, since the boundary condition
arises from assuming an infinite mass for the fermion outside the bag. 

For $m=0$ the solutions to Eqs. (\ref{eq:lfbag}) read
\bea
\psi_{+} &=& e^{-i{\cal E} \zeta} \nonumber\\
\psi_{-} &=& const.
\eea
and thus from the boundary conditions (\ref{eq:lfbc})
\be
{\cal E}_n = \left( n+\frac{1}{2} \right) \frac{2\pi}{L}.
\ee
Using again ${\cal E}_n = {2} E_n$, we find that the spectrum obtained
in tilted LF coordinates is consistent with Eq. (\ref{eq:bagEn}).

\subsection{Three Dimensional Harmonic Oscillator}

In order to solve the Helmholtz equation in tilted coordinates
\begin{equation}
{\cal P}{\cal E} \phi = 
   \left[m^2+\kappa
      \left({\vec x}_\perp^2+\zeta^2\right)
      +{\vec p}_\perp^2 
      +{\cal P}^2\right]  \phi,
\label{eq:HOtilt}  
\end{equation}
we first complete the square, yielding
\be
{{\cal E}^2\over 4}\phi =  \left({\cal P} - \frac{{\cal E}}{2}\right)^2\phi
+  \left[m^2+\kappa \left({\vec x}_\perp^2+
\zeta^2\right)+
{\vec p}_\perp^2\right]\phi .
\label{eq:HOtilt2}
\ee
First of all, we note that a shift in ${\cal P}$, included by multiplying the
wave function $\phi$ by a suitably chosen phase factor can be made to absorb
the dependence of the r.h.s. of Eq. (\ref{eq:HOtilt2}) on ${\cal E}$. 
Furthermore, the
transverse and longitudinal dynamics in Eq.(\ref{eq:HOtilt2}) completely
separates, yielding
\be
{{\cal E}^2_n\over 4} =  m^2 + \omega_L \left(n_L+\frac{1}{2}\right)
+ \omega_\perp \left(n_\perp+1\right),
\label{eq:HOtilt3}
\ee
where $\omega_L= \omega_\perp = 4 \kappa=\omega$. Taking into account 
that ${\cal E}_n = {2}E_n$ we thus find that the relativistic 
harmonic oscillator formulated in an equal time 
framework and in tilted LF coordinates yield identical spectra.
\subsection{Dirac Equation with Static Potential}
We express Eq.~(\ref{eq:dirac_lf1}) in tilted coordinates to find:
\begin{eqnarray}
& &\left[\left(i\gamma^+{\partial\over\partial \tau} - i\gamma^3
{\partial\over \partial \zeta}\right)
+i{\vec \gamma}_\perp\cdot\left( {\vec \partial}_\perp
  + g_v {\vec \partial}_\perp
  \Gamma({\vec r}_\perp, -\zeta) \right)- m \right.
\label{eq:tdirac_lf1}
\\
& &\quad \left.- 
V_S({\vec r}_\perp, -\zeta)- 
\gamma^+V_0 ({\vec r}_\perp, -\zeta )
\right] \psi^\prime =0. \nonumber\end{eqnarray}
The potentials $V_S$ 
and $V_0$ are functions of $r_\perp^2+\zeta^2$, so we expect rotational  
invariance to be immediate. Indeed, the above is just the standard Dirac 
equation, except for the appearance of $\gamma^+$ instead of $\gamma^0$.
This can immediately be removed by using the transformation of 
Eq.~(3.44) of Ref.\cite{us}. Thus the spectra of the Dirac equation in tilted
 coordinates is  going to be the same as that of the usual  equal time 
formulation.

\section{Summary}

The three features discussed in the introduction emerge naturally from
a more general notion that  static sources
in the usual ET formulation become static sources in the LF formulation
if one makes an appropriate transformation (\ref{eq:tilde}). The resulting
LF formulation for scalar interactions can be summarized as simply including a
$p^+$ in the $x^+$ development
operator, and evaluating the potentials using the
replacement $z\to -x^-/2$. The resulting simplicity allows us solve any problem
involving a spherically symmetric static source. An application of this in
Sect.~V.D enables us to obtain a momentum distribution which increases it
width as a function of $p_\perp$ for increasing values of $p^+$.
A hint of such behavior has recently been observed\cite{ep}. 

We note that the spectrum condition--the feature that light front mode
functions have support only for positive values of
plus-momentum-- has not been 
maintained in any of the solutions we present here.  The desire to 
maintain this 
consistency with the
original field theory  led us  to implement a new numerical
procedure in Ref.~\cite{us}. In the case of 
the harmonic oscillator, we see that for the general
conditions of nuclear physics, the spectrum condition is maintained, even though
the solution procedure ignores this condition. 
The detailed study of the 
how important maintaining the spectrum
condition, for models other than the harmonic oscillator,
 is will be a topic of future investigation.
The numerical results of Ref.~(\cite{us}), for nucleons in nuclei,
 strongly indicate   that there is no strong need to maintain the spectrum
 condition. The presence of a 5\% effect for $q(x=0)$ indicates that the 
maintaining the spectrum condition is important for massless quarks.  

The simplicity of the results for the simple modeled considered, along with the
need to automate the transformation procedures, 
 suggests the use of a new set of coordinates--
tilted coordinates in which the $p^+$ term is in from the beginning and in
which static sources in the ET formulation are also static sources in the
tilted LF formulation. For the case of the simple models considered here,
this new formulation is actually easier to apply
than the usual LF formulation. Whether or not this simplicity  survives more
detailed problems is a matter for future investigation.
\section*{Acknowledgments}
This work is partially supported by the USDOE. We thank B.~Shlaer for useful
discussions.

\end{document}